\documentstyle[prc,eqsecnum,aps,epsfig]{revtex}
\tightenlines
\begin{document}
%
%
\title{Production of intermediate-mass dileptons
in relativistic heavy ion collisions}

\author{Ioulia Kvasnikova and Charles Gale}

\address{Department of Physics, McGill University,\\
3600 University Street, Montr\'eal, QC, H3A 2T8 Canada}


\author{Dinesh Kumar Srivastava\thanks{Permanent address: Variable Energy 
Cyclotron Centre, 1/AF Bidhan Nagar, Kolkata 700 064, India}}

\address{Department of Physics, Duke University, Durham, NC 27705}

\date{\today}

\maketitle 
\begin{abstract}
The production of intermediate mass dileptons in
ultrarelativistic nuclear collisions at SPS energies is studied. 
The acceptance and detector resolution inherent to measurements by the NA50
experimental collaboration are accurately modeled.  
The measured centrality dependence of the intermediate mass lepton 
pair excess is also addressed. 
\end{abstract}

\pacs{ }

\section{Introduction}

The study of relativistic heavy ion collisions derives its unique
importance from the opportunity of producing hot and dense strongly 
interacting matter. More specifically, one of the hopes is the  
creation and study of a quark-gluon plasma (QGP), a prediction of QCD
\cite{kars}, 
which filled the nascent Universe roughly a microsecond after the Big Bang.
This field of endeavour currently represents a cutting edge of 
research in subatomic physics
and has generated tremendous activity, both in theory and in
experiment \cite{qm01}. Good penetrating probes of in-medium phenomena in heavy
ion collisions are electromagnetic observables, owing to their small 
rescattering cross sections and to their privileged coupling to vector
mesons \cite{sakurai}.  For collisions at ultrarelativistic energies, the
lepton pair  invariant mass (M) spectrum can roughly be divided in 
three regions: low mass (M $< m_\phi$), high mass (M $> m_{J/\psi}$),
and intermediate mass. The high mass region is Drell-Yan-dominated at
the CERN SPS, and
this contribution is in principle calculable within the realm of perturbative QCD.
In connection with the low mass sector in heavy ion collisions 
at similar energies,  the  measurements
of the CERES collaboration \cite{ceres} have been
seminal in building the case for an observation of in-medium effects on
the vector meson spectral densities \cite{rw}. There, the experimental
collaboration has identified an excess over the sources that were known
to be sufficient for the understanding of the spectra in pA
collisions. This modification of the vector meson properties in matter
represents an exciting discovery and is a genuine consequence of  
many-body physics. 

In the intermediate invariant mass region, an excess of dimuons  over the 
sources expected from pA measurements has also been identified 
experimentally by the Helios/3 \cite{helios3} and NA50 \cite{na50ejpc} 
collaborations. Some excitement and interest  
have also been associated with this observation as this mass region is
the one originally predicted to contain a plasma signal \cite{first}.
Theoretical models have been devised to understand this dimuon
excess. Those include rescattering of the open charm mesons
with the surrounding hadronic constituents \cite{lw98}.  This approach
can not however reproduce the dimuon mass and transverse momentum 
spectra for central nuclear collisions \cite{capelli}. An overall charm
production enhancement has also been invoked as a prospective interpretation
\cite{na50ejpc}. Whereas this scenario raises an interesting 
possibility, it has to be reconciled 
quantitatively with the observation of $J/\psi$ suppression.
This issue can however be settled by a direct measurement \cite{na60}.   
An earlier theoretical study of the Helios/3 measurements has attributed the
excess to secondary hadronic reaction processes \cite{gali}. This
result has some consequences that are important to put on a firmer
basis, by comparing with other newer data. In this strain, more recent
NA50 results have also been interpreted by thermal models, with some
quark-gluon content \cite{RS,kampfer}. 

Our goal in this paper is to consider the NA50 intermediate mass dimuon
results in the light of a hydrodynamic modeling of the nuclear
dynamics, of a detailed analysis of the hadronic dilepton rates, and of
a precise simulation of the detector acceptance and resolution. 
The issue of centrality dependence of the signal is also addressed. 
The next section briefly describes the hydrodynamic approach utilized,  
along with its reproduction of hadron spectra. The 
dilepton rate calculation is then described and the numerical
modeling of the detector cuts and resolution is introduced. After
the dilepton spectra in invariant mass and transverse momentum have been
discussed, the centrality dependence is considered. The paper closes 
with a summary and a conclusion.

\section{Hydrodynamic evolution and hadron spectra}

It is well known that the final yield of the thermal dileptons has to
be obtained by an integration over the space-time history of evolution
of the interacting system. We shall assume that a thermally and chemically
equilibrated quark-gluon plasma is produced in such collisions at a time
$\tau_0$, which may be estimated from the condition of isentropic
expansion of the plasma~\cite{bj}:
\begin{equation}
\frac{2\pi^4}{45\zeta(3)}\,\frac{1}{A_T}\frac{dN}{dy}=4 a
T_0^3\tau_0
\end{equation}
where $dN/dy$ is the particle rapidity density for the collision,
and $a=42.25 \pi^2/90$ for a plasma of massless u, d, s quarks, and gluons.
The number of flavours is taken to be $N_f\approx 2.5$ to account
for the mass of strange quarks.
We see that once the transverse area of the colliding system $A_T$ is
known, along with $dN/dy$, the above relation uniquely relates $T_0$
to $\tau_0$. While one may get arbitrary large value of $T_0$ by choosing
a smaller formation time of the plasma, the uncertainty 
principle
provides a lower physical limit on $\tau_0 \sim 1/3T_0$~\cite{kms}.

While considering Pb + Pb collisions at SPS energies, we have
taken the  average particle rapidity density as 750 for the 10\% most
 central Pb + Pb collisions at the CERN SPS energy, 
as measured.  We estimate the average number of
participants for the corresponding range of impact parameters
($0\, \leq\, b \, \leq \, 4.5$ fm) as  about 380, compared to the maximum of
416 for  a head-on collision. We thus use a mass number of 190 to get the
 radius of the
transverse area of the colliding system and neglect
its deviations from azimuthal symmetry, for simplicity.
As the deviation measured in terms of the number of participants is
marginal ($<$ 9\%), we expect the error involved also to be small. We also
recall that the azimuthal flow is minimal for
near
 central collisions.

The plasma is then assumed to undergo a boost-invariant longitudinal
expansion and an azimuthally symmetric radial expansion, with a 
transition to a hot hadronic gas consisting of {\em all} hadrons 
having M$ < $ 2.5 GeV, in a thermal an chemical equilibrium at
a transition temperature $T_c$. This makes for a rich equation of
state. Once all parton matter is converted 
into hadronic matter, we assume the hot hadronic system to continue to
expand until it undergoes a freeze-out at some temperature $T_F$. During
this evolution, the speed of sound of the matter is consistently calculated
at every temperature to be used in the equation of state needed for solving
the hydrodynamic equations~\cite{crs}.

It is important to realize that the (pressure) gradients in the system
cause and control the radial expansion of the plasma, which in turn
leads to a more rapid cooling and to the development of a transverse 
velocity for the fluid. Support for these ideas comes from intensity
interferometry  and from measured particle spectra.
It is thereby necessary to have a proper initial energy density profile
as it affects the hydrodynamic developments by introducing 
additional gradients.
We assume it to follow the so-called
``wounded-nucleon'' distribution, which for central collision of identical
nuclei leads to:
\begin{equation}
\epsilon(\tau_0,r)\propto \int_{-\infty}^\infty \rho(\sqrt{r^2+z^2})\, dz
\end{equation}
where $\rho$ is the (Woods-Saxon) distribution of nucleons 
and $r$ is the transverse distance. This is prompted by
the experimental observation that transverse energy deposited in these
collisions scales with the number of participants.  The normalization is 
then 
determined through numerical integration such that
\begin{equation}
A_T \, \epsilon_0=\int \, 2\pi \, r \,  \epsilon(r) \,\,dr.
\end{equation}

A proper profile is also needed to get a quantitative description of the
dilepton yields and through this to infer the initial conditions.
 It is clear that the energy densities at larger $r$ would
be smaller, implying lower temperatures and a shorter life-time
till the fluid there undergoes freeze-out. Now recall that the four-volume
$d^4x=\tau \, d\tau \,r\,dr\, d\eta\, d\phi$ and 
thus the contribution of a given
fluid element rises linearly with its transverse distance, which 
further ``magnifies'' the contribution from large $r$.
One can immediately see that a uniform energy density profile often
employed in the literature would lead to an 
erroneously
 large contribution
from large $r$. This fact plays an important role
in arriving at the required initial conditions. 

We now have all the ingredients to solve the hydrodynamic equations.
This is done through the procedures described in Refs.~\cite{crs,hydro}.

\subsection{Initial conditions at the SPS}
On the basis of the rapidity density etc. observed at SPS energies
in central collisions involving lead nuclei, it was shown 
recently~\cite{dks_wa98}
 that
a good empirical description of the single photon data 
measured by the WA98 collaboration was obtained when the plasma 
formation time was taken to be $\tau_0=$ 0.2 fm/$c$. This 
corresponded to an average initial temperature of about 330 MeV. In the
present work, we shall use these values but also explore the consequences
of varying $\tau_0$ and $T_0$, keeping the resulting $dN/dy$ fixed.

The phase transition is assumed to  take place at $T=$ 180 MeV and the
freeze-out at 120 MeV. This value of the critical temperature
is motivated by lattice QCD results which give values of about
170 -- 190 MeV~\cite{kars}, and the thermal model analyses of hadronic
ratios which
suggest that the chemical freeze-out in such collisions takes place
at about 170 MeV (recent analyses yield a value
 of $158.1\,\pm\,3.2$ MeV~\cite{johanna} for the chemical freeze-out
temperature). 
The phase transition should necessarily take place at a higher
temperature.

\subsection{Hadronic Spectra}

As a first step we look at the spectra of several hadrons for the
central collision of lead nuclei measured at the SPS. 
\begin{figure}
\begin{center}
\includegraphics[width=8.2cm]{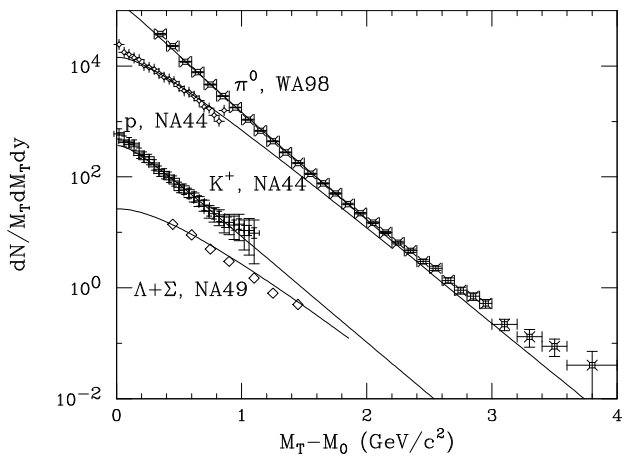}
\end{center}
\caption{ Hadronic spectra in central collision of lead nuclei at
SPS energy. The formation time is assumed to be 0.2 fm/$c$. The data are
for pions, protons, kaons, and lambdas + sigmas, from top to bottom, 
respectively.}
\label{hadspect1}
\end{figure}
\begin{figure}
\begin{center}
\includegraphics[width=6cm,angle=90]{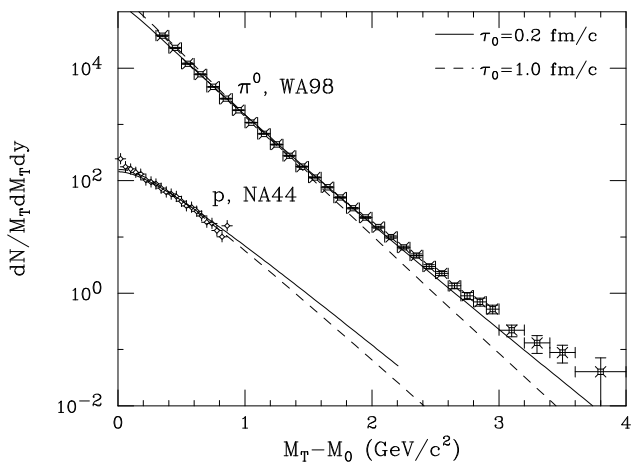}
\end{center}
\caption{ Hadronic spectra in central collision of lead nuclei at
SPS energy. The formation time is taken as 0.2 fm/$c$ and 1 fm/$c$
respectively.}
\label{hadspect2}
\end{figure}
In Fig.~\ref{hadspect1} 
we show our results when the formation time is take as $\tau_0=$
0.2 fm/$c$ as in Ref.~\cite{dks_wa98}. The data are from Refs.
\cite{wa98,na44,na49}.  Even though it is known
for quite some time that  a reasonable variation of the initial
temperature and the formation time, keeping the corresponding
$dN/dy$ fixed, affects the flow only marginally, we study the
consequences of increasing $\tau_0$ to 1 fm/$c$ in Fig.~\ref{hadspect2}. 
We see that
the hadronic spectra shapes are relatively insensitive to such variations in the 
initial conditions, as the flow takes some time to develop. In all cases, the 
agreement is quite good. This fact lends some credibility to the dynamics
advocated here.

\section{Dilepton Spectra at SPS energy}

\subsection{Dilepton emission rates}

For the dilepton emission from the deconfined sector of QCD, we have assumed
that the dominating process is $q \bar{q} \to \ell^+ \ell^-$. The
emission rates that correspond to this process have been calculated 
many times. An example of their quantitative contribution can be found
in Ref.~\cite{k2m2}. Even though those rates are the Born terms in the
parton sector, a very recent lattice calculation has found that those were
surprisingly accurate \cite{karsch2}. 

\begin{figure}
\begin{center}
\includegraphics[width=7cm]{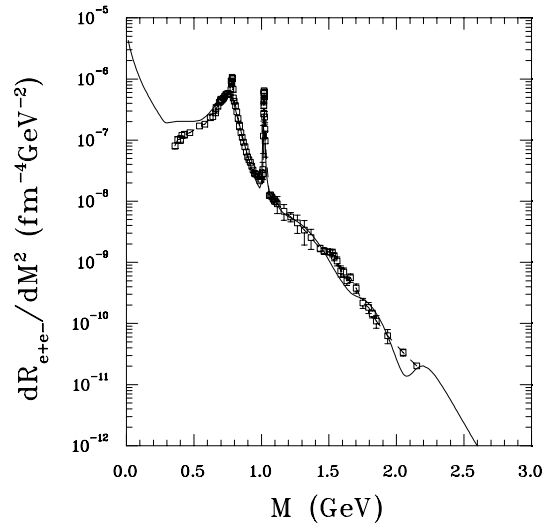}
\end{center}
\caption{Net dilepton production rate from a gas of mesons at a
temperature of T = 150 MeV, as a
function of dilepton invariant mass. The full curve is the sum of the
hadronic channels discussed in the text and in the references.  The
data points follow from an extraction of spectral densities from
$e^+ e^-$ scattering experiments \protect\cite{huang,huang2}.}
\label{hadrate}
\end{figure}

In the evaluation of hadronic rates most calculations have relied on
effective Lagrangian techniques, where the available parameters
are fitted to empirical measurements. An example of the
application of this line of thought  to the low mass dilepton sector 
can be found in Refs.~\cite{rw,ragale}. A problem which arises when one
tries to extend those techniques to the intermediate mass region is the
appearance of off-shell effects. 
Indeed, different approaches that agree in the soft sector
can yield widely different results in higher invariant mass extrapolations
\cite{gaogale}. Fortunately, constraints on the hadronic processes can
be obtained through the wealth of data of the type $e^+ e^- \to$
hadrons \cite{dol91}. In addition, those measurements cover exactly
the same invariant mass range as the one which concerns this work.  
Those have been used, together with $\tau$-decay data, to
construct the vector and axial-vector spectral densities \cite{huang}. 
Similarly, the intermediate invariant mass initial state $e^+ e^-$ data
has been analyzed specifically in the most important exclusive channels.
This information can then be used to construct rates for hadrons $\to
e^+ e^-$ \cite{gali}. It is this procedure that is followed in the
current work.  The contributing channels producing lepton pairs 
in the invariant mass range
1 GeV $<$ M $<$ 3 GeV have been found to correspond to the initial
states: $\pi \pi$, $\pi \rho$, $\pi \omega$, $\eta \rho$, 
$\rho \rho$, $\pi a_1$, 
$K \bar{K}$, $K \bar{K^*} + c.c.$ \cite{gali,ioulia}. A detailed
channel-by-channel discussion is too long to be had here, but a 
quantitative
assessment of the net dilepton  rate from a gas of mesons at T = 150
MeV is shown in Fig.~\ref{hadrate}. Also shown is the rate extracted
from the spectral function evaluation \cite{huang,huang2}. In
principle, the latter contains all the hadronic sources.  
The correspondence between those two results is a confirmation that the
important channels have indeed been identified in the hadronic
scenario. 

\begin{figure}
\begin{center}
\includegraphics[width=8.2cm]{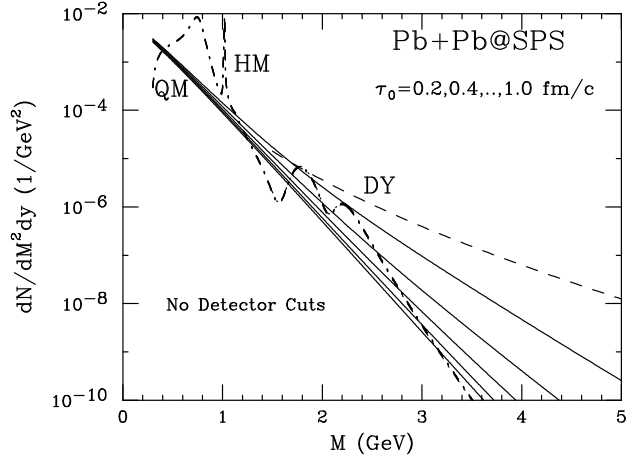}
\end{center}
\caption{The net yield of lepton pairs from the quark gluon plasma,
Drell-Yan and the hadronic sources are shown separately, before
detector and acceptance corrections. The QGP signal for different
formation times $\tau_0$ are shown (solid  lines), earlier time on top, later times
on the bottom. The Drell-Yan contribution is the dashed line, and the hadronic
matter contribution the dashed-dotted curve. The yield from correlated open
charm decay is not shown here. That source's contribution to dN/dM is a 
decaying exponential in the mass range under scrutiny and its dependence on
various parameter assumptions is discussed in detail in \protect\cite{capelli}.}
\label{fignocuts}
\end{figure}

Using the rates of dilepton production from quark and hadronic matter, 
it is straightforward to calculate the spectra for dileptons, say
for the central collision of lead nuclei at SPS energy. 
We add the contributions of the quark matter 
from the QGP phase and the quark matter part of the mixed phase and call
it QM. Similarly we add the contributions of the hadronic matter part of
the mixed phase and the hadronic phase and call it HM. The contribution
of the Drell-Yan process is obtained by a scaling
of nucleon-nucleon estimates to the case of nucleus-nucleus collisions.
We get the results shown in Fig.~\ref{fignocuts}, no acceptance or
resolution corrections are implemented.
It is clear that in principle, 
if the formation time is small ({\it i.e.}, the initial temperature is large)
then there could be a substantial contribution of the dileptons having their 
origin in the quark matter, in the intermediate mass window. 
Note that the HM contribution is only very marginally affected by
variations in the $\tau_0$. We shall proceed to model the effect of detector
resolution and to compare with experimental measurements. 


\subsection{Detector acceptance and dilepton spectra} 

It is vital to account for the finite acceptance of the detectors and
for their resolution when comparing the results of theoretical
calculations with measured experimental data. Those effects are
important in the NA50 experiment \cite{capelli}.  One approach to this
problem is to model approximately and analytically the acceptance. 
While this can be readily implemented \cite{RS,iouqm}, a legitimate doubt 
can subsist about the accuracy of the experimental representation,
especially in regions where edge effects might be important. In order
to circumvent this problem we use here for the first time a numerical
subroutine developed to reproduce the NA50 acceptance cuts and 
finite resolution effects in  the measurements of dimuon pairs in Pb + Pb
collisions at the CERN SPS \cite{drapier}. 

\begin{figure}
\begin{center}
\includegraphics[width=8.2cm]{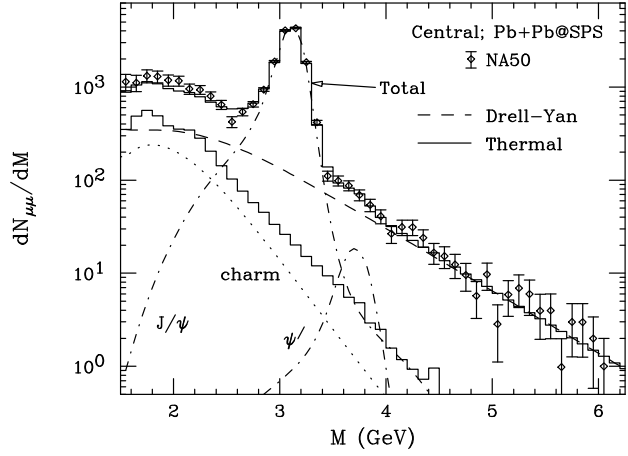}
\end{center}
\caption{We show the calculated dimuon invariant mass 
spectrum after correcting for detector acceptance and resolution. 
The data is
from the NA50 collaboration \protect\cite{na50ejpc}. The Drell-Yan and thermal
contributions are shown separately, as well as those coming from
correlated charm decay and direct decays of the $J/\psi$ and $\psi^\prime$.}
\label{fig:dsigdm}
\end{figure}

\begin{figure}
\begin{center}
\includegraphics[width=8.2cm]{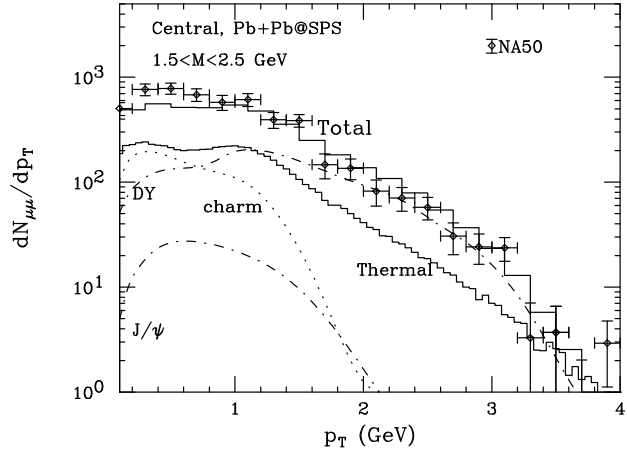}
\end{center}
\caption{We show the dimuon transverse momentum
spectrum after accounting for detector effects. 
The origin of the data and of the different sources are as in
Fig.~\protect\ref{fig:dsigdm}.} 
\label{fig:pt}
\end{figure}

\begin{figure}
\begin{center}
\includegraphics[width=8.2cm]{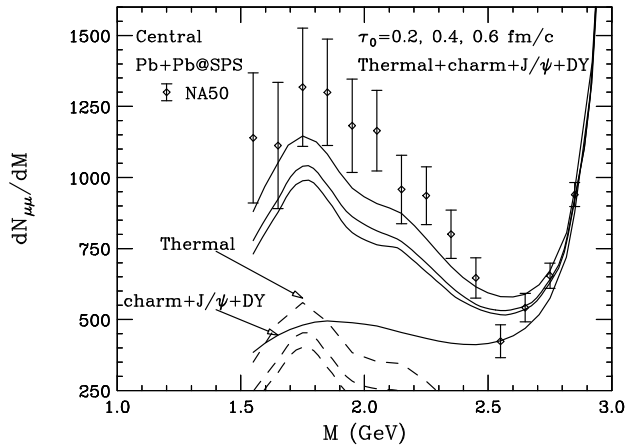}
\end{center}
\caption{A linear plot of the net dilepton spectrum in the
intermediate mass region. The three solid curves correspond to
formation times $\tau_0$ = 0.2, 0.4, 0.6 fm/c, from top to bottom,
respectively. The data are from \protect\cite{na50ejpc}. The thermal
contributions and the contribution from hard processes are shown
separately.}
\label{linear}
\end{figure}

We compute the invariant mass distribution of the dileptons in our
hydrodynamic approach and then we run our pairs through the numerical 
detector simulation. The normalization is determined by a fit to
the Drell-Yan data using the MRSA parton distribution functions as in the
NA50 analysis. For getting the $p_T$ distribution we supplement our
$dN/dM^2$ estimates for the Drell-Yan with a gaussian distribution
in $p_T$ as in Ref.~\cite{RS} which very closely reproduced the
estimates obtained by the NA50 collaboration.
The result is shown in Fig.~\ref{fig:dsigdm}. We
also compute the $p_T$ distribution for the muon pairs, in the
invariant mass window 1.5 $<$ M $<$ 2.5 GeV. This spectrum is broken
down in the same sources as previously and is shown in
Fig.~\ref{fig:pt}. 
At this point it is appropriate to consider the following question:
which initial temperature is demanded by the intermediate mass
dilepton data? A critical and quantitative assessment of this issue
can be obtained by examining a linear plot of the lepton pair spectrum in
the mass region under scrutiny. This is show in Fig.~\ref{linear}. 
From this figure, it appears that the best fit is provided by $\tau_0$
= 0.2 fm/c, and that the second best (less than two standard
deviations away for most of the data points) belongs to $\tau_0$ = 0.4
fm/c. In terms of initial temperature those two values of the formation time
correspond to $T_0 \approx$ 330 and 265 MeV respectively. A
conservative and reasonable  point of view is that it is
probably not fair in such a challenging and complex environment as that of
ultrarelativistic heavy ion collisions to ask for an agreement
that is better than two standard deviations, considering the inherent
uncertainties. This circumvents the issue of initial temperature 
determination. Also, there are questions
that have to do with the specific dynamics used here to model the
nuclear collisions. However in this respect, the hydrodynamic  model used in
this work 
is solved self-consistently and its parameters are constrained by 
the hadronic data as well. In connection with the important discussion 
on plasma signals 
the quark matter contribution is $\approx$ 23\% for $\tau_0$ = 0.2
fm/c, and $\approx$ 19\% for $\tau_0$ = 0.4 fm/c, 
around the lepton pair invariant mass of 1.5 GeV. 

\subsection{Centrality dependence}

\begin{figure}
\begin{center}
\includegraphics[width=6cm,angle=90]{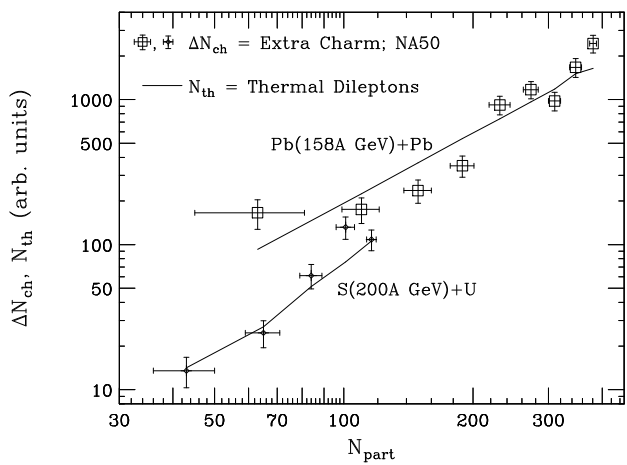}
\end{center}
\caption{Centrality dependence. The data represent the ``extra charm''
yield (as categorized by the NA50 collaboration
\protect\cite{na50ejpc}) needed to describe the intermediate mass
region dimuon data. The full curve is from our thermal sources.}
\label{fig:central}
\end{figure}

Our discussion so far was concerned only with the high multiplicity
bin or, with mostly central collisions. To extend the hydrodynamic
model to non-central events and to properly treat the azimuthal
anisotropy is not a simple task. However, one can get a fair estimate
of the
centrality  dependence by ignoring the broken azimuthal
symmetry and by approximating the region of nuclear overlap by a circle
of radius $R \approx 1.2 (N_{\rm part} / 2)^{1/3}$, where $N_{\rm
part}$ is the number of participants~\cite{scale}. Assuming the uncertainty
relationship, $\tau_0 = 1/3 T_0$, one can track the changing multiplicities 
without introducing additional parameters.  The results of this exercise are
shown in Fig.~\ref{fig:central}, for S + U and Pb + Pb collisions. 
Also shown are the  contributions from the ``enhanced production
of charm'' estimated by the NA50 collaboration, normalized to the 
dileptons estimated by us.
We estimate the
contribution of the ``excess charm'' $\Delta N_{ch}$ from
\begin{equation}
\Delta N_{ch} \propto (E-1)N_{coll}
\end{equation}
where $N_{coll}$ is the average number of collisions for the given 
centrality and $E$ is the `enhancement' factor given by the
NA50 measurement \cite{capelli}.
 The constant of proportionality 
is identical for all the bins for a given system. 
It is seen that the agreement with the measured data is quite good,
and our approach gives a fair description of the centrality dependence of
the excess dilepton measurement.

\section{Summary and conclusion}

In the light of the findings reported on in this work, it does appear
that the case of attributing the intermediate mass dilepton excess to an
enhanced charm production rests on evidence that is no longer very
compelling. The
lepton pairs coming from a thermal hadronic medium comprise an
important part of the signal, in agreement with earlier analyses
of the Helios/3 measurements \cite{gali}, and with those of more 
recent studies \cite{RS,kampfer}. 
It is
satisfying to notice that the behaviour of the centrality dependence 
is also reproduced by the modeling and the dynamics used here. 
We have also considered the constraints on the initial temperature set by the
data in the framework of our approach.
There is a QGP component in the
treatment used in this work. Unfortunately, its numerical value is not
important enough to provide irrefutable evidence. 
The future of this field is bright: RHIC 
electromagnetic measurements 
should soon signal bold incursions into a new territory.

\acknowledgments

It is a pleasure to acknowledge useful conversations with L. Capelli, 
O. Drapier, and L. Kluberg. 
This work was supported in part by the Natural Sciences and Engineering
Research Council of Canada, in part by FCAR of the Quebec
government, and in part by the US Department of Energy under grant
DE-FG02-96ER40945..


\bigskip
 
\begin{thebibliography}{99}
\bibitem{kars}F. Karsch, {\it 17$^{th}$ International Symposium on
Lattice Field Theory} (Lattice '99), Pisa, Italy, hep-lat/9909006.

\bibitem{qm01}See, for example, {\it Proceedings of the 15th.
International Conference on Ultrarelativistic Nucleus-Nucleus
Collisions (Quark Matter 2001)}, Nucl. Phys. {\bf A} in press, and
references therein. 
\bibitem{sakurai}J. J. Sakurai, {\it Currents and Mesons}, (University
of Chicago Press, Chicago 1969); H. B. O'Connell, B. C. Pearse, A. W.
Thomas, and A. G. Williams, Prog. Part. Nucl. Phys. {\bf 39}, 201
(1997).
\bibitem{ceres}H. Appelshaeuser, {\it Proceedings of the 15th.
International Conference on Ultrarelativistic Nucleus-Nucleus
Collisions (Quark Matter 2001)}, Nucl. Phys. {\bf A} in press, and
references therein. 
\bibitem{rw}R. Rapp and J. Wambach, Adv. Nucl. Phys. {\bf 25}, 1 (2000).  
\bibitem{helios3}M. Masera {\it et al.}, Nucl. Phys. {\bf A590}, 93c
(1995); A. L. S. Angelis {\it et al.}, Eur. J. Phys. C {\bf 13}, 433
(2000). 
\bibitem{na50ejpc}M. C. Abreu {\it et al.}, NA50 Coll.
 Eur. Phys. J. C {\bf 14}, 443
(2000). 
\bibitem{first} 
                E. V. Shuryak, Phys. Lett. {\bf 78}, 150 (1978).
\bibitem{lw98}Z. Lin and X. N. Wang, Phys. Lett. {\bf B444}, 245
(1998). 
\bibitem{capelli}L. Capelli, Ph. D. thesis, Universit\'e Claude Bernard
(2001).
\bibitem{na60} http://na61.web.cern.ch/NA6i/
\bibitem{gali}G. Q. Li and C. Gale, Phys. Rev. Lett. {\bf 81}, 1572 (1998); 
Phys. Rev. C {\bf 58}, 2914 (1998). 
\bibitem{RS}R. Rapp and E. Shuryak, Phys. Lett. {\bf B473}, 13 (2000). 
\bibitem{kampfer}K. Gallmeister {\it et al.}, Nucl. Phys. {\bf A688}, 939
(2001). 
\bibitem{bj} J. D. Bjorken, Phys. Rev. D {\bf 27}, 140 (1983);
              R. C.Hwa and K. Kajantie, Phys. Rev. D {\bf 32}, 1109 (1985).
\bibitem{kms} J. Kapusta, L.McLerran, and D. K. Srivastava,
              Phys. Lett. B {\bf 283}, 145 (1992).

\bibitem{crs}
J. Cleymans, K. Redlich, D.K. Srivastava,  
Phys. Rev.C {\bf 55}, 1431 (1997);
J. Cleymans, K. Redlich, D. K. Srivastava;
Phys. Lett. B {\bf 420}, 261 (1998). 

\bibitem{hydro} H. von Gersdorff, L. McLerran, M. Kataja, and P. V.
Ruuskanen, Phys. Rev. D {\bf 34}, 794 (1986); P. V. Ruuskanen, Acta
Phys. Pol. B {\bf 18}, 551 (1986).

\bibitem{dks_wa98} D. K. Srivastava and B. Sinha, Phys. Rev. C {\bf 64},
034902 (2001).
\bibitem{johanna} 
 See, e.g., P. Braun-Munzinger, I. Heppe, and J. Stachel, 
Phys. Lett. B {\bf 465}, 15 (1999) See, e.g., P. Braun-Munzinger, I. Heppe, and J. Stachel, 
Phys. Lett. B {\bf 465}, 15 (1999);
 F. Becattini, J. Cleymans, A. Keranen, E Suhonen, and K. Redlich,
 Phys. Rev. C {\bf 64}, 024901 (2001). 
\bibitem{wa98}M. M. Aggarwal {\it et al.}, Phys. Rev. Lett. {\bf 81},
4087 (1998).
\bibitem{na44}I. G. Bearden {\it et al.}, Phys. Rev. Lett. {\bf 78},
2080 (1997).
\bibitem{na49}P. G. Jones {\it et al.}, Nucl. Phys. {\bf A610}, 175c
(1996).
\bibitem{k2m2} K. Kajantie, J. Kapusta, L. McLerran, and A. Mekjian, 
Phys. Rev. D {\bf 34}, 2746 (1986). 
\bibitem{karsch2}F. Karsch {\it et al.}, hep-lat/0110208.
\bibitem{ragale}Ralf Rapp and Charles Gale, Phys. Rev. C {\bf 60}, 024903
(1999). 
\bibitem{gaogale}Song Gao and Charles Gale, Phys. Rev. C {\bf 57}, 254 (1998).  


\bibitem{dol91}See, for example, S. I Dolinsky {\it et al.}. Phys. Rep.
{\bf 202}, 99 (1991), and references therein.
\bibitem{huang}Z. Huang, Phys. Lett. {\bf B361}, 131 (1995). 

\bibitem{ioulia}I. Kvasnikova, Ph. D. thesis, McGill University (2001).
\bibitem{huang2}Z. Huang, private communication.
\bibitem{iouqm}Dinesh Kumar Srivastava, Bikash Sinha, Ioulia Kvasnikova, 
and Charles Gale, {\it Proceedings of the 15th.
International Conference on Ultrarelativistic Nucleus-Nucleus
Collisions (Quark Matter 2001)}, Nucl. Phys. {\bf A} in press. 

\bibitem{drapier}O. Drapier, for the NA50 collaboration, private
communication; NA38 collaboration, Nucl. Instrum. Methods, {\bf A405},
139 (1998). 

\bibitem{scale} D. K. Srivastava, Phy. Rev. C {\bf 64}, 064901 (2001);
J.-Y. Ollitrault, Phys. Lett. B {\bf 273}, 31 (1991).




















\end{thebibliography}
\end{document}